\documentclass[a4paper,11pt]{article}
\pdfoutput=1
\usepackage{jheppub}
\usepackage{amsmath}
\usepackage{amsthm}
\usepackage{amsfonts}
\usepackage{amssymb}
\usepackage{textcomp}
\usepackage{graphicx}
\usepackage{bm}
\usepackage{color}
\usepackage{verbatim}
\usepackage{subcaption}
\usepackage{graphicx,color}
\usepackage{epsfig}
\usepackage{caption}

\usepackage{amsfonts,amsmath,amssymb}

\makeatletter

\@addtoreset{equation}{section}
\makeatother

\begin{document}

\begin{titlepage}

\vfill


 \title{Gravitational quasinormal modes for Lifshitz black branes}

 \author[a]{Tomas Andrade}
 \author[b]{Christiana Pantelidou} 
 
 \affiliation[a]{Departament de Fisica Quantica i Astrofisica \& Institut de Ciencies del Cosmos Universitat de Barcelona (ICCUB), 
 Marti i Franques 1, 08028 Barcelona, Spain} 
 \affiliation[b]{School of Mathematics and Statistics, University College Dublin, Belfield, Dublin 4, Ireland} 
 
\abstract{We study the scalar and vector channels of gravitational quasinormal modes for Lifshitz black branes emerging in Einstein-Maxwell-Dilaton and Einstein-Proca theories in four and five dimensions, finding significant differences between the two models. In particular, rather surprisingly, in the Einstein-Maxwell-Dilaton  model the dispersion relations for the shear and sound modes are given by $\omega_{shear} \sim -i\,k^4$ and $\omega_{sound}\sim-i \,k^2$, while in the Einstein-Proca model they take the more conventional form  $\omega_{shear} \sim -i\,k^2$ and $\omega_{sound}\sim k$ \textemdash the proportionality constants depend on the dynamical exponent and the appropriate factors of temperature. Through the holographic duality, this calculation  provides information about the relaxation of the momentum and energy flux operators in a putative dual Lifshitz field theory. Comparing with the dispersion relations obtained directly by considering Lifshitz hydrodynamics suggest that the mass density of the equilibrium state in the Einstein-Maxwell-Dilaton model is infinite.} 
 
\maketitle

\end{titlepage}

\section{Introduction}

Black holes, unlike many idealised physical systems, are intrinsically dissipative due to the presence of an event horizon. Thus, when considering the characteristic oscillations of these systems,  instead of carrying out a standard normal-mode analysis, one reverts to the computation of quasinormal modes (QNMs). The latter have in principle complex frequencies, with the real part representing the actual frequency of the oscillation and the imaginary part representing the damping.

QNMs have been studied in various contexts. On the one hand, in asymptotically flat spacetimes they have been used in the context of black hole spectroscopy to infer the mass and angular momentum of the final black hole created after a binary merger as well as for testing no-hair theorems. On the other hand, in asymptotically $AdS$ spacetimes, QNMs of black branes have been used for studying the near-equilibrium behaviour of strongly coupled plasmas with a dual gravity description revealing intriguing connections between the dynamics of horizons and relativistic hydrodynamics.  For a recent review on QNMs for asymptotically flat and asymptotically $AdS$ black holes/branes see \cite{Berti:2009kk}.

In this paper we study QNMs in black brane geometries that asymptote to the so-called Lifshitz geometry\footnote{Note that here we are focusing our attention to spatially isotropic Lifshitz geometries. Non-spatially isotropic Lifshitz geometries were obtained in \cite{Taylor:2008tg}.} described by the line element (for a review see \cite{Taylor:2015glc}).
\begin{align}
ds^2&=-r^{2z} dt^2+\frac{dr^2}{r^2}+r^2 dx_idx^i\,,
\end{align}
where $z$ is the critical exponent and $i=1,\dots ,D-2$. These geometries manifestly realise the Lifshitz symmetry $Lif_{D-2}(z)$ (rather than relativistic invariance) which comprises of temporal ($H$) and spatial ($P^i$) translations, spatial rotations ($L^{ij}$) as well as a scaling symmetry $\mathcal{D}_z$
\begin{align}
H:& \quad t\to t'=t+a\,\nonumber\\
P^i:&\quad x^i\to {x^\prime}^{i}=x^i+a^i\,\nonumber\\
 L^{ij}:&\quad x^i\to {x^\prime}^{i}=L^i_{\,j}x^j\nonumber\\
\mathcal{D}_z: &\quad r\to r'=\lambda r,\quad t\to t'=\lambda t,\quad x^i\to {x^\prime}^{i}=\lambda^z x^i\,.
\end{align}
When $z = 1$ the metric is $AdS$ and has full relativistic symmetry, but for $z\ne1$ the system obtains anisotropic scaling between space and time and is thus non-relativistic. For $z\ge 1$, Lifshitz geometries satisfy the  strong energy condition $R_{mn}u^m u^n\ge0$ for any future directed timelike vector $u^m$, as well as the null energy condition $G_{mn}k^m k^n\ge0$ for any future directed null vector $k^m$.  Therefore there are no obstruction to supporting the Lifshitz geometry with physically reasonable matter for $z\ge 1$. For $z<1$, some pathologies emerge caused by the violation of the null energy condition \cite{Hoyos:2010at}.

Lifshitz geometries have an anisotropic curvature tensor and solve the Einstein equations with a non-trivial stress energy tensor. They were first constructed in \cite{Kachru:2008yh} and have been mainly realised in Einstein-Maxwell-Dilaton (EMD) \cite{Taylor:2008tg,Way:2012gr}, Einstein-Proca (EP)\cite{Taylor:2008tg} and higher derivative gravity theories\cite{Taylor:2008tg,Cai:2009ac,Ayon-Beato:2010vyw,Matulich:2011ct}. Lifshitz black holes have been constructed in various dimensions and for different values of $z$, with some analytic and many numerical examples, see e.g. \cite{Ayon-Beato:2019kmz, Bravo-Gaete:2020ftn, Bravo-Gaete:2021kgt}.

According to the holographic duality, these geometries are dual to strongly-coupled non-relativistic field theories with Lifshitz symmetry. However, the corresponding holographic dictionary is less well-understood than in the case of $AdS$ and in fact a real-time formulation is still not fully developed. Holographic renormalisation, which is what allows one to determine the independent sources and their corresponding expectation values, is more elaborate \cite{Ross:2009ar,Ross:2011gu,Griffin:2011xs}. 
These references find that the operator content of the dual theory corresponds to a non-relativistic stress tensor complex, consisting of an energy density ${\cal E}$, energy flux ${\cal E}_i$, momentum density ${\cal P}_i$ and spatial stress-tensor $\pi^{ij}$. In the absence of sources, these satisfy the usual Ward identities
\begin{align}
	\partial_t {\cal E} + \partial_i {\cal E}^i &= 0 \\
	\partial_t {\cal E}^i + \partial_j \pi^{ij} & =0 
\end{align}
In addition, a common feature present in holographic realizations of Lifshitz is the appearance of a scalar operator with irrational scaling dimensions. This can be associated to the longitudinal polarization of the massive vector in EP, or the dilaton in EMD theories. 

Recent studies of a specific $z=2$ Lifshitz geometry arising as reduction of higher-dimensional theories\cite{Christensen:2013rfa} suggested that Lifshitz field theories couple generically to Newton-Cartan geometries on the boundary \textemdash note that $z=2$ is special as the symmetry group can be augmented by Galilean boosts. This study has lead to some resurgence of interest in non-relativistic holography, including studies of intrinsically non-relativistic gravity theories such as Horava-Lifshitz \textemdash this is to be contrasted with what has been stated above, where one considers relativistic gravity theories that admit non-relativistic solutions.

Scalar field QNMs in Lifshitz black brane backgrounds were considered previously in the literature. In particular, the cases that have been studied include $D = 3, z = 3$ in NewMassiveGravity \cite{Cuadros-Melgar:2011wen}, $D\ge 4, z=2$ in $R^2$ gravity \cite{Abdalla:2012si} and $D\ge 2, z=2$ in a $R^3$ gravity \cite{Giacomini:2012hg}. Additionally, the case of $D= 4 , z = 2$ has been studied in the Einstein-Proca-Scalar theory \cite{Gonzalez:2012de}, in the EMD setup \cite{Myung:2012cb} and in a topological black hole in a Einstein-Maxwell-Proca background \cite{Gonzalez:2012xc,Mann:2009yx,Brynjolfsson:2009ct,Becar:2015kpa}. All these quasinormal modes were found to be purely imaginary. This is in contrast with the results of \cite{Sybesma:2015oha} for the EMD model that showed that, at zero momentum, these modes have purely imaginary frequencies for $z\ge D-2$, while for $z< D-2$ they pick up a real part. In all case, all quasinormal modes are situated in the lower half-plane of complex frequencies, indicating stability.
Furthermore, in the context of holographic superconductors in a Lifshitz background, \cite{Natsuume:2018yrg} found hydrodynamic modes in the QNM spectrum of the charged scalar field close to the critical temperature.

In this work we study gravitational QNMs, focusing on the EP and the EMD model for $D=4,5$ and general $z$. These QNMs are more interesting than scalar ones because the corresponding fluctuations couple to conserved currents in the dual field theory. Currently, these have only been calculated for the EP model for $D=4, z=2$ at zero temperature \cite{Zingg:2013xla} and negative imaginary frequencies were found. Our results for the leading behaviour of the dispersion relations of the gapless (hydrodynamic) modes are summarised in the table below

\begin{center}
\begin{tabular}{ l|l } 
\hspace{2cm} EP & \hspace{1.3cm} EMD \\ 
  \hline
$\omega_{shear}=-i\,\nu(z) k^2+\dots$ & $\omega_{shear}= -i\,\bar{\nu}(z) \,k^4+\dots$ \\ 
 \hline
$\omega_{sound}=u_s(z) \, k-i \Gamma(z) k^2+\dots$ & $\omega_{sound}= -i \bar{\Gamma}(z)\,k^2\dots$  \\ 
\end{tabular}
\end{center}
\vspace{0.5cm}

\noindent where $\omega, k\ll1$ are the frequency and the momentum of the modes respectively. Note that all the $z$ dependence is hidden in the constant of proportionality that we have calculated numerically. We see substantial differences in the relaxation of the two theories, and in the case of the EMD theory, we also see significant deviation from the dispersion relations of the hydrodynamic modes in $AdS_D$, which take the form 
\begin{equation}
\omega_{shear}= -i\, \nu\,k^2+\dots\,,\qquad \omega_{sound}= u_s k-i \Gamma k^2+\dots
\end{equation}
where $u_s$ is the speed of sound, $\Gamma$ is the attenuation and $\nu$ is the diffusion constant. In addition to the above hydrodynamic modes, we also find  a tower of non-hydrodynamic modes the real parts of which go to zero at $z = D-2$, just like for the scalar QNMs \cite{Sybesma:2015oha}.
All the quasinormal modes found are located in the lower half plane and we thus conclude that the system is stable, which is in agreement with the non-linear time evolution of \cite{Gursoy:2016tgf} for the EMD model.

Through the holographic duality, QNMs correspond to poles of the (retarded) thermal correlators of dual $(D-1)$-dimensional strongly interacting quantum field theories. The lowest QNM frequencies of black branes have a direct interpretation as dispersion relations of hydrodynamic excitations in the dual field theory, which in our case enjoys Lifshitz symmetry. Lifshitz hydrodynamics have been developed in \cite{Hoyos:2013eza,Hoyos:2013qna} and more recently in \cite{deBoer:2017abi}. This has been carried out in two competing approaches: both start with relativistic hydrodynamics but break the Lorentz symmetry in different ways along the hydrodynamic expansion. In particular, the difference lies on whether Galilean boosts are broken at the perfect fluid level or at the first dissipative order. In both cases new transport coefficients were identified and the bulk viscosity is found to vanish\cite{Hoyos:2013eza,deBoer:2017abi}.  Dispersion relations for the hydrodynamic sound, shear and diffusion modes have been studied in \cite{deBoer:2017abi} and a new expression for the speed of sound has been obtained\cite{deBoer:2017abi,Hoyos:2013cba}. The sound mode dispersion relation in a higher-derivative gravity theory for $z=3$, $D=3$ at finite temperature in the hydrodynamic limit was also studied in \cite{Bhattacharyya:2015bsa}.

The remaining on this paper is structured as follows. In section \ref{sec:QNM} we give a brief overview of the computation of gravitational QNMs and in sections \ref{sec:EP} and  \ref{sec:EMD} we discuss, respectively, EP and the EMD theories and the corresponding results that we obtained. Then in section \ref{sec:FT} we carry out a comparison with the Lifshitz hydrodynamics developed in \cite{deBoer:2017abi} and finally in section \ref{sec:discussion} we conclude with some discussion and future directions.

\section{Review of quasinormal mode computation} \label{sec:QNM}
  
A detailed analysis of the quasinormal spectra for AdS-Schwarzschild and AdS-Reissner-Nordstrom black branes was discussed previously in the literature \cite{Kovtun:2005ev}. 
It is well known that the electromagnetic and gravitational perturbations split into the tensor (for $D \ge 5$),  vector and scalar sectors depending on their transformation properties. 
The scalar sector contains the sound and charge diffusion fluctuations, the vector contains the shear and transverse gauge field fluctuations and the tensor modes are decoupled scalar equations. 
Out of these fluctuations only the shear, sound and charge diffusion contain hydrodynamic modes, meaning modes that obey dispersion relations such that the frequency approaches zero as the momentum is decreased. These hydrodynamic modes correspond precisely to the shear and sound modes in the hydrodynamic limit of the dual CFT at finite $T$ we are interested in.

In the case of  black branes with Lifshitz asymptotics, the electromagnetic and gravitational fluctuations once again split into the three sectors described above. In this work we focus only on the scalar and vector fluctuation, which contain gapless (hydrodynamic) modes, meaning modes that obey dispersion relations such that the frequency approaches zero as the momentum is decreased. Note that in our case, even though we have a vector field in the bulk, there is no U(1) current in the dual theory, so only sound and shear modes associated to the stress tensor are expected. In particular, we do not expect to find charge diffusion hydrodynamic mode in the set-ups we are considering.

In particular, we consider linearised fluctuations around the Lifshitz backgrounds of the form
\begin{align}
&g_{\mu\nu}=g^{(Lifz)}_{\mu\nu}+\epsilon\, \delta g_{\mu\nu}(r)\, e^{-i(\omega t-k x)}+\dots\,\nonumber\\
&A_{\mu}= A^{(Lifz)}_\mu+\epsilon\, \delta A_\mu(r)\,e^{-i(\omega t-k x)}+\dots\,\nonumber\\
&\delta \phi=\phi^{(Lifz)}+\epsilon \,\delta\phi(r)\, e^{-i(\omega t-k x)}+\dots\, \text{(for the EMD model)}
\end{align}
where $\epsilon$ is a small expansion parameter. Note that here we have chosen the  momentum $k$ to point in the $x$ direction without lose of generality. The precise form of the Lifshitz background, denoted by $g^{(Lifz)}, A^{(Lifz)}$ and $\phi^{(Lifz)}$ is described in detail in the sections below.

In the vector sector, the non-trivial metric and gauge field fluctuations considered are
\begin{equation}
\{\delta g_{ti}, \delta g_{xi},\delta A_i,\delta g_{ri}\}\,,
\end{equation}
where $i$ denotes the spatial directions transverse to $x$, in which we retain isotropic e.g. $\delta g_{xy}=\delta g_{xz}$ in $D=5$. 

In the scalar sector, the non-trivial metric and gauge field fluctuations considered are
\begin{equation}
\{\delta g_{tt},, \delta g_{tx},\delta g_{xx}, \delta g, \delta A_t, \delta A_x,\delta g_{tr}, \delta g_{rr}, \delta g_{rx}, \delta A_r\}\,,
\end{equation}
where $\delta g=\frac{1}{D-3}\sum_i\delta g_{ii}$. 

The equations of motion for these fluctuations carry a lot of redundant information due to gauge invariance. In particular, under an infinitesimal coordinate transformation $x^\mu\to x^\mu+\xi^\mu$, where $\xi^\mu$ is an arbitrary function of $r$, the fluctuations transform as
\begin{align}
\delta g_{\mu\nu}&\to\delta g_{\mu\nu}-\nabla_\mu\xi_\nu -\nabla_\nu\xi_\mu\,,\nonumber\\
\delta A_\mu&\to \delta A_\mu+\nabla_\mu\lambda-\xi^\nu \nabla_\nu A_\mu-A_\nu\nabla_\mu \xi^\nu\,,\nonumber\\
\delta\phi&\to\delta\phi-\xi^\nu\nabla_\nu \phi\,.
\end{align}
Typically this gauge freedom is dealt with by arranging the fluctuations into gauge invariant combinations \cite{Kovtun:2005ev}; this is the approach we follow in the majority of our calculations. For the scalar channel of the Einstein-Proca model we instead follow a method first discussed in \cite{Rangamani:2015hka} that mimics the DeTurck trick\footnote{This method was also used for double checking some of our results for the sound channel in the Einstein-Maxwell-Dilaton model}. Specifically, we add a gauge fixing term $\nabla_{(\nu}\tau_{\mu)}$ to Einstein's equations, where
\begin{align}
\tau_\mu=\nabla^\nu\left( \delta g_{\nu\mu}- det(\delta g) \frac{g_{\nu\mu}^{(Lifz)}}{2}\right)\,,
\end{align}
 and a posteriori check that it vanishes. Note that such term is only added for the metric and not for the gauge field. Since the gauge field is massive, the equation of motion completely determines $A$, which means that one can not do any gauge transformations. Comparing the two approaches, considering gauge invariant combinations is preferred as it boils down to a smaller number of equations to be solved numerically.

The final equations are then solved numerically subject to boundary conditions, namely ingoing boundary conditions at the horizon and fast enough fall-off close to the UV boundary compatible with the absence of sources. 

\section{Einstein-Proca theory}\label{sec:EP}

 \subsection{The model}
In this section we consider the Einstein-Proca model, described by the following bulk action
 \begin{equation}
 S= \int dx^{D}\sqrt{-g}\left( R- 2 \Lambda-\frac{1}{4} F_{\mu\nu}F^{\mu\nu}-\frac{m^2}{2} A_\mu A^\mu\right) \,,
\end{equation}
where $F=dA$. The corresponding equations of motion are given by
\begin{align}
R_{\mu\nu}&= \frac{2 \Lambda}{3} g_{\mu\nu}+\frac{1}{2} F_{\mu\lambda}F_{\nu}^{\,\,\lambda}
+\frac{m^2}{2}A_\mu A_\nu-\frac{1}{12} F_{\rho\sigma}F^{\rho\sigma}g_{\mu\nu}\,,\nonumber\\
\nabla_\mu F^{\mu\nu}&= m^2 A^\nu\,.
\end{align}
and it is easy to show that for 
\begin{align}
\Lambda=-\frac{z^2+(D-3)z+(D-2)^2}{2}\,,\quad m^2=(D-2)z\,,
\end{align}
they admit a solution
\begin{align}
ds^2&=r^2(-r^{2(z-2)}dt^2+dr^2+dx^idx^i)\,,\nonumber\\
A&=r^z\sqrt{\frac{2(z-1)}{z}}dt\,,
\end{align}
which corresponds to a dual field theory at zero temperature with Lifshitz symmetry.

In order to put this system at finite temperature, one needs to consider configurations with regular horizons that approach the above the solution close to the boundary. This was achieved in \cite{Way:2012gr}, using the ansatz
\begin{align}\label{eq:LifEP}
ds^2&=r^2\left(-r^{2(z-1)}F(r)dt^2+\frac{dr^2}{R(r)}+dx^idx^i\right)\,,\nonumber\\
A&=r^z\sqrt{\frac{2(z-1)}{z}} a_t(r)dt\,,
\end{align}
which gives rise to two second order equations for $F, a_t$ and an algebraic equation for $R$. These equations can be solved numerically using a shooting method subject to appropriate UV boundary conditions and regularity at the horizon (located at $r=r_h$). In particular, close to the boundary ($r\to\infty$), the fields fall off according to
\begin{align}
F&=1+  c_1 r^{-D+2-z}+c_+ r^{-\lambda_+/2}+c_- r^{-\lambda_-/2}+\dots\,,\nonumber\\
a_t&=1+  c_2 c_1 r^{-D+2-z}+ c_3 c_+ \lambda_+ r^{-\lambda_+/2}+c_4 c_- \lambda_-r^{-\lambda_-/2}+\dots\,,
\end{align}
where $c_1,c_\pm$ are undetermined constants, $c_2,c_3,c_4$ are functions of $z$ (which we omit for simplicity) and
\begin{align}
\lambda_\pm=\begin{cases}
		z+3\pm\sqrt{9 z^2-26z+33} \qquad \text{for D=5}\nonumber\\
		z+2\pm\sqrt{9 z^2-20z+20} \qquad \text{for D=4}\,.
		\end{cases}
\end{align}
In our UV boundary conditions we require that $c_-=0$, following the analysis of \cite{Ross:2009ar}.  The temperature of the solution is finite and given by
\begin{equation}
T=\frac{r_h^{z+1}}{4\pi}\sqrt{\frac{R}{F}} F'\large|_{r=r_h}\,.
\end{equation}

\begin{figure}[h!]
\centering
\includegraphics[width=0.5\linewidth]{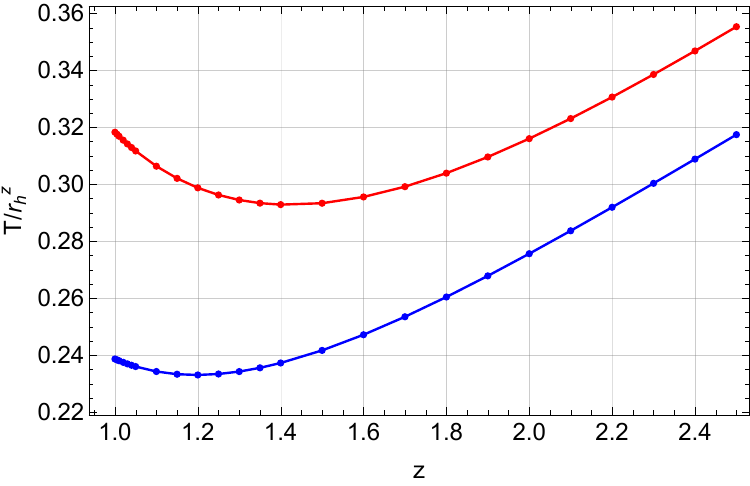}
\caption{Temperature as a function of $z$ for Lifshitz black branes in the Einstein-Proca model for $D=4$ (blue) and $D=5$ (red) . Note that the curves approach the AdS values $T_{D=4} = 3 r_h/(4 \pi)$, $T_{D=5} = r_h/\pi$ at $z = 1$.}
\label{fig:EP_bkg}
\end{figure}

In Fig \ref{fig:EP_bkg} we plot the temperature dependence with $z$ for the numerical solutions we have constructed.  

 \subsection{Numerical results}

\subsubsection{Vector channel}
\begin{figure}[h!]
\centering
\includegraphics[width=0.48\linewidth]{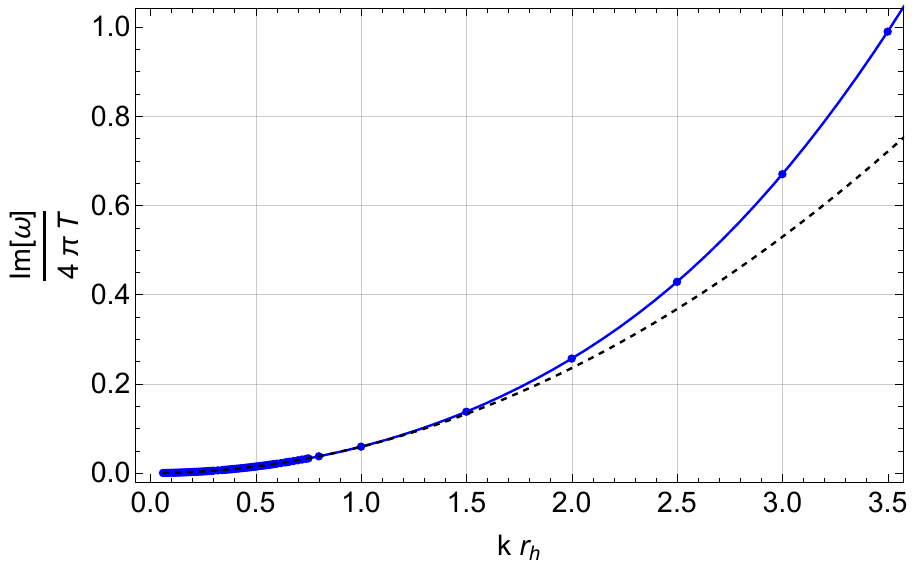} \quad \includegraphics[width=0.48\linewidth]{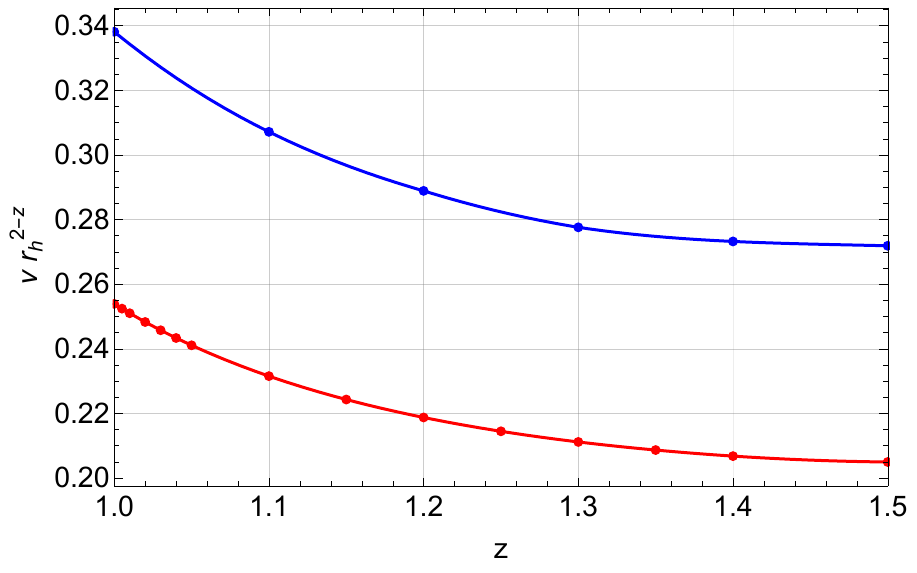}
\caption{Hydrodynamics modes for the shear channel in the EP model. \textbf{(a)} Plot of the dispersion relation for the shear mode for $z=1.2$ for the $D=5$ EP model. \textbf{(b)} Values of the diffusion constant $\nu(z)$ for $D=4$ (blue) and $D=5$ (red).}
\label{fig:shearEP}
\end{figure}

Given the vector channel fluctuations, the equations of motion reduce to a set of two second order ODEs for the gauge invariant quantities 
\begin{align}
\delta A_y, \quad \delta H_{xy}=r^4\partial_r(r^{-2} \delta g_{xy})\,,
\end{align}
 which we omit for simplicity.  The independent terms in the boundary asymptotics for the gauge invariants can be written as
\begin{align}
&\delta A_y=a^{(0)} r^z+a^{(1)} r^{-2z+4-D},\qquad \delta H_{xy}=r^{3-2z}H^{(0)} +r^{-z-D+5}H^{(1)}\,.
\end{align}
It is easy to see that setting the sources to zero requires $a^{(0)} = H^{(0)} = 0$ \cite{Andrade:2013wsa}. At the horizon we require ingoing boundary conditions, which implies that the fields behave as
\begin{equation}
\delta A_y=(r-r_h)^{-i\omega/4\pi T} a_y^{(reg)}\,,\qquad \delta H_{xy}=(r-r_h)^{-i\omega/4\pi T-1}  \delta H_{xy}^{(reg)}\,,
\end{equation}
where $T$ is the temperature and $a_y^{(reg)}, H_{xy}^{(reg)}$ admit regular power series expansions in the near horizon region. We determine the spectrum of QNMs by discretising the differential equations and solving the corresponding matrix problem in Mathematica.  We find one gapless mode 
\begin{equation}
\omega_{shear}=-i \nu(z) k^2+\dots
\end{equation}
In Fig. \ref{fig:shearEP}(a) we plot of the dispersions relation for the shear mode for $z=1.2$ for $D=5$, corresponding to a clear quadratic power-law behaviour for small momenta. Similar plots have been obtained for different values of $z$ and for $D=4$, when perturbing around the corresponding numerical backgrounds. The diffusion constant depends on the value of $z$  and in particular it decreases  as $z$ is increased, as shown in Fig.  \ref{fig:shearEP}(b).
 
\subsubsection{Scalar channel}
We now consider the scalar channel fluctuations around the Lifshitz background \eqref{eq:LifEP}. As explained in the previous section, we follow the gauge fixing approach of \cite{Rangamani:2015hka}, which requires introducing an extra term in the equations of motion for the metric \textemdash we have verified that this term vanishes within numerical precision for the modes of interest, so that these fluctuations are indeed solutions of the linearised Einstein's equations. 

\begin{figure}[h!]
\centering
\includegraphics[width=0.48\linewidth]{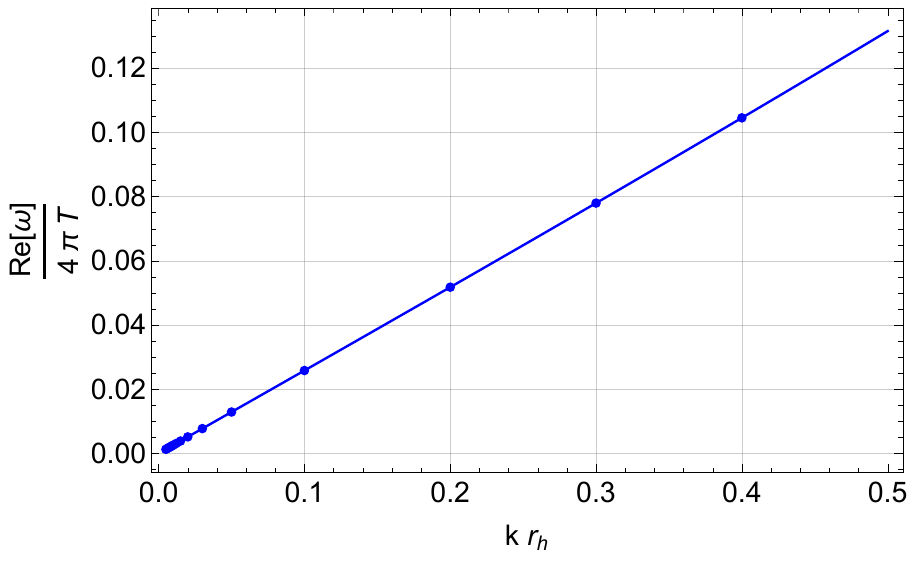}\quad \includegraphics[width=0.48\linewidth]{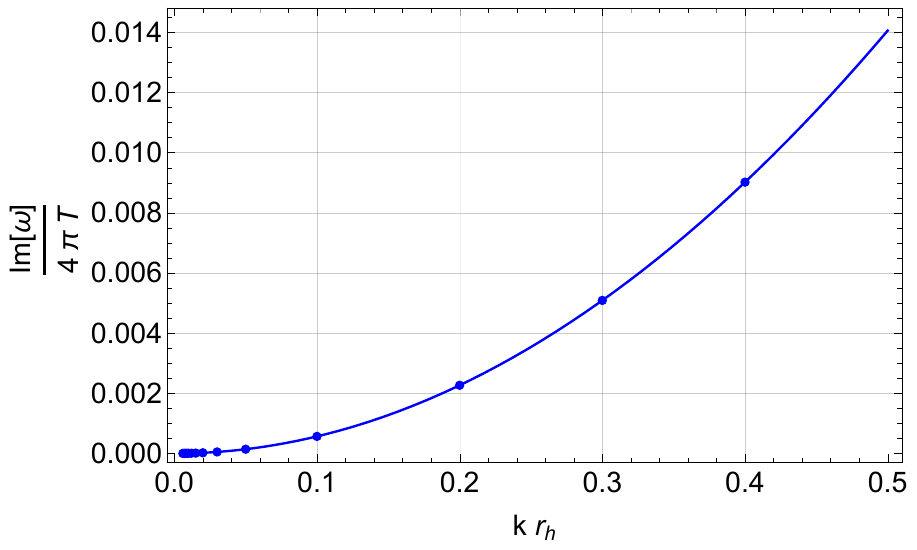}
\caption{Hydrodynamics modes for the sound channel in the EP model. {\bf(a)} real and {\bf(b)} imaginary part of the dispersion relation for $z=1.5$ and $D=4$.}
\label{fig:soundEP}
\end{figure}

\begin{figure}[h!]
\centering
\includegraphics[width=0.48\linewidth]{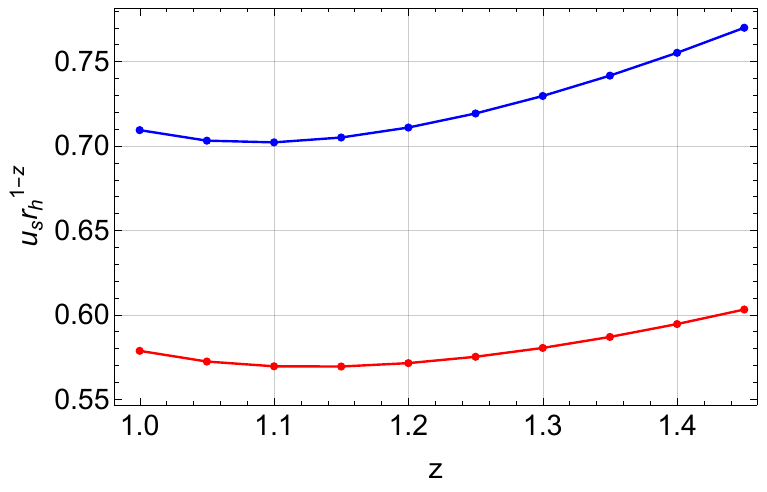}\quad \includegraphics[width=0.48\linewidth]{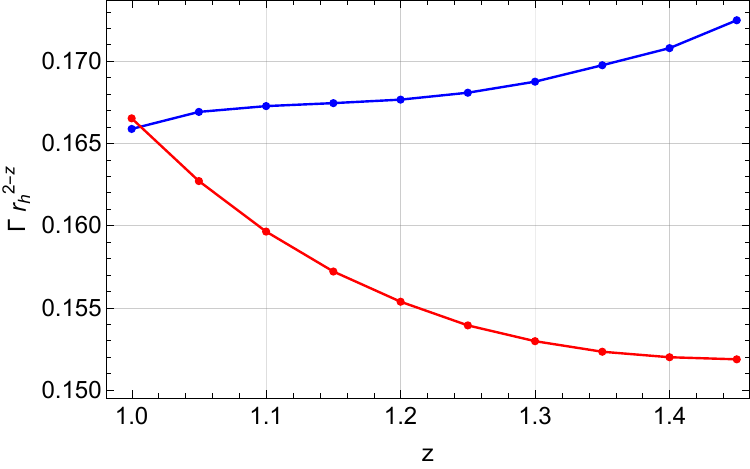}
\caption{Coefficients of the dispersion relation for the sound channel in the EP model as a function of $z$. {\bf(a)} $u_s$ and {\bf(b)} $\Gamma$ for $D=4$ (blue) and $D=5$ (red).}
\label{fig:dispersionEPvsz}
\end{figure}

The equations of motion for these fluctuations are solved numerically by discretisation subject to ingoing boundary conditions at the horizon and fast enough fall-off close to the boundary compatible with the absence of sources; see appendix \ref{app:EP} for more details. We find one gapless mode characterised by the standard dispersion relation
\begin{equation}
\omega_{sound}= u_s(z) k - i \Gamma(z) k^2 + \dots
\end{equation}
\noindent We show in Fig \ref{fig:soundEP} the behaviour of $\omega(k)$ for $z=3/2, D=4$; we find analogous behaviour 
for other values. We display the numerical values of $u_s(z)$ and $\Gamma(z)$ in Fig. \ref{fig:dispersionEPvsz}. Note that our results agree with the AdS values in e.g. \cite{Morgan:2009pn}. 

\vspace{0.5cm}


\section{Einstein-Maxwell-Dilaton theory}\label{sec:EMD}
 \subsection{The model}
 In this section we consider the theory described by the bulk action
 \begin{equation}
 S= \int dx^{D}\sqrt{-g} \left(R-2\Lambda +\frac{1}{4} e^{\lambda \Phi} F_{\mu\nu}F^{\mu\nu}-\frac{1}{2}\partial_\mu \Phi\partial^\mu \Phi\right)\,,
 \end{equation}
 which gives rise to the following equations of motions
\begin{align}
&R_{\mu\nu}-\frac{1}{2} e^{\lambda\Phi}F_{\mu\rho}F_{\nu}^{\,\,\rho}-\frac{1}{2}\partial_\mu \Phi\partial_\nu \Phi
+ g_{\mu\nu}\left(-\frac{2\Lambda}{d-2}+\frac{1}{4(d-2)}e^{\lambda \Phi}F_{\mu\nu}F^{\mu\nu}\right)\,,\nonumber\\
&\nabla_\mu \left(e^{\lambda\Phi}F^{\mu\nu}\right)- m^2 A^\nu=0\,,\nonumber\\
&\nabla^2 \Phi-\frac{\lambda}{4}e^{\lambda\Phi}F_{\mu\nu}F^{\mu\nu}=0\,.
\end{align}
  As it was shown in \cite{Taylor:2008tg} (for generalisations see \cite{Tarrio:2011de}), this theory admits an analytic solution black hole configuration given by
  \begin{align}\label{eq:LifEMD}
ds^2&=-r^{2z} f(r)dt^2+\frac{dr^2}{r^2 f(r)}+r^2 dx_idx^i\,,\nonumber\\
e^{\lambda\Phi}&=e^{\lambda\Phi(r)}=r^{-2(D-2)}\,,\quad A=a_0 r^{z+D-2}\,,\quad f(r)=1-\frac{r_h}{r}^{z+D-2}\,,
 \end{align}
 where
 \begin{align}
 \Lambda=-\frac{1}{2}(z+D-2)(z+D-3)\,,\quad \lambda^2=2\frac{D-2}{z-1}\,,\quad a_0=\sqrt{\frac{2(z-1)}{\lambda^2(z+D-2)}}\,.
 \end{align}

 It is easy to see that close to the boundary $r\to \infty$ this solution approaches the Lifshitz metric with critical exponent $z$, while the black hole horizon is located at $r=r_h$ corresponding to temperature
  \begin{equation}
  T=\frac{z+D-2}{4\pi} r_h^z\,,
 \end{equation}
 where $r_h$ is the radius of the horizon and in our units $r_h=1$.
 
\noindent It is worth mentioning that these configurations are not smoothly connected to AdS branes in the limit $z \to 1$. This is easily seen by noting the divergence in the coupling $\lambda$ which appears explicitly in the equation of motion for the dilaton.  
 
 \subsection{Numerical results}
 
 \subsubsection{Vector channel}
\begin{figure}[h!]
\centering
\includegraphics[width=0.49\linewidth]{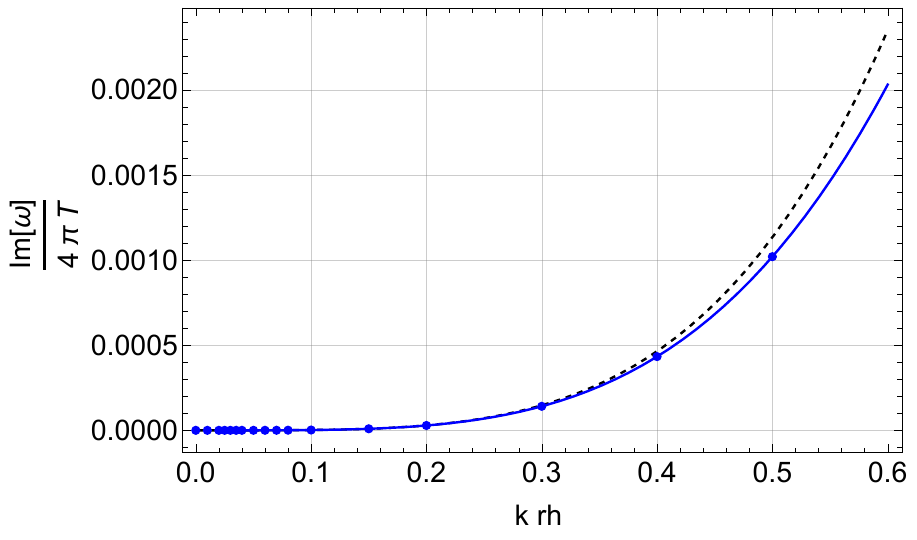}\quad\includegraphics[width=0.48\linewidth]{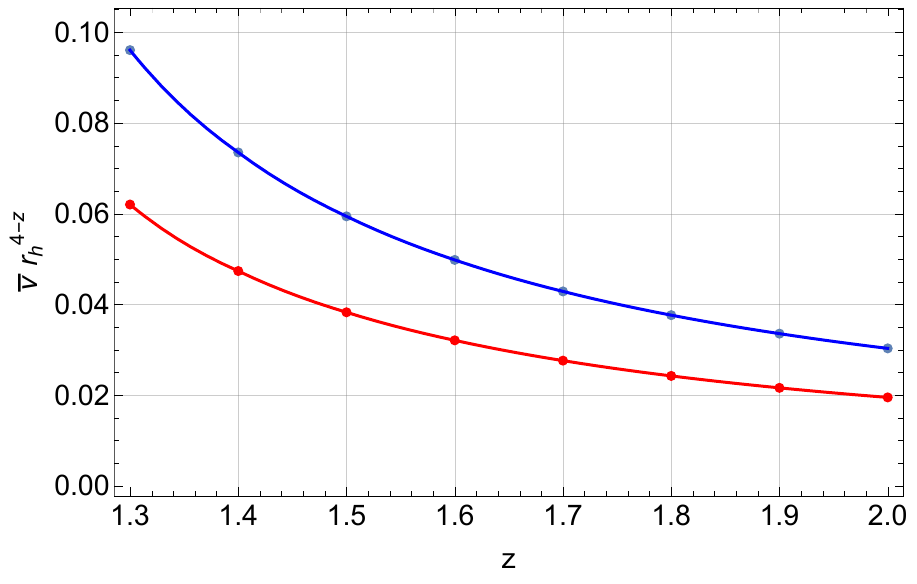}
\caption{Hydrodynamics modes for the shear channel in the EMD model. {\bf (a)} Plot of the dispersion relation for $D=4$ and $z = 3/2$. {\bf (b)} Plots of $\bar \nu(z) $ as a function of $z$ for $D=4$ (blue) and $D=5$ (red).}
\label{fig:barGamma}
\end{figure}

We consider the vector channel fluctuations around the background \eqref{eq:LifEMD} in the gauge where $\delta g_{ri}=0$ and we focus our attention on the gauge invariant quantities 
\begin{equation}
\delta A_y,\quad \delta H_{ty}=a_0 r_h^{D-z+1} \partial_r(r^{-2} \delta g_{ty})\,,
\end{equation}
 which obey second order ODEs.  The independent terms in the boundary asymptotics for the gauge invariants can be written as
\begin{align}
&\delta A_y=a^{(0)} r^{D-2+z}+a^{(1)}r^{2-2z}\,,\quad  \delta H_{ty}=r^{-3+2z}H^{(0)} +r^{1-D-z}H^{(1)}\,.
\end{align}
The leading terms parametrise the field theory sources so we set them to zero. Given these boundary conditions in the UV and ingoing boundary conditions at the horizon,
\begin{equation}
\delta A_y=(r-r_h)^{-i\omega/4\pi T} a_y^{(reg)}\,,\qquad \delta H_{ty}=(r-r_h)^{-i\omega/4\pi T} \delta H_{ty}^{(reg)}\,,
\end{equation}
 we proceed to solve the ODEs numerically to find the spectrum of QNMs. We find one gapless mode with dispersion
\begin{equation}
\omega_{shear}=-i \bar \nu(z) k^4+\dots
\end{equation}
In Fig \ref{fig:barGamma}(a) show the dispersion relation for $D=4$, $z=3/2$, which behaves line $\sim k^4$ in the hydrodynamic limit \textemdash \,this was also confirmed by studying the logarithmic derivative of the dispersion $k Im[\omega]'/Im[\omega]$. We find an analogous behaviour for $D = 5$. The dispersion constant, $\bar\nu$, depends on the value of $z$ in a way which we depict in Fig \ref{fig:barGamma}(b). 

  \subsubsection{Scalar channel}
  In this section we consider the scalar sector fluctuations around the background \eqref{eq:LifEMD}. In particular, in the gauge where $\delta g_{tr}$, $\delta g_{rr}$, $\delta g_{rx}$, $\delta a_r=0$, these fluctuations are combined into the following gauge invariant combinations
  \begin{align}
&  Z_1=\delta\phi - \frac{1}{2r(d-2)} \phi' \left(2 \delta g_{xx}+(D-4) \delta g\right)-\frac{2}{\lambda}\frac{1}{r^{2}}\left(\delta g_{xx}-\delta g\right)\,,\nonumber\\
&  Z_2=\delta a_t+\frac{\omega}{q}\delta a_x- \frac{1}{2r(D-2)} a_t' \left(2 \delta g_{xx}+(d-4) \delta g\right)\,,\nonumber\\
&  Z_3=r^{2z-2} f\delta g_{tt}- 2\frac{\omega}{q} \frac{1}{r^{2}}\delta g_{tx}-\frac{\omega^2}{q^2}( \delta g_{xx}- \delta g)-\frac{1}{2r} (r^{2z} f)' \delta g\,.
  \end{align}
 The equations of motion for these master fields take the form of three second order ODEs of degree 6 in the frequency. At the horizon we impose ingoing boundary conditions
  \begin{equation}
 Z_i=(r-r_h)^{-i\omega/4\pi T} Z_i^{(reg)}\,, \qquad i=1,2,3
 \end{equation}
where $ Z_i^{(reg)}$ admit regular power series expansions.  A mode analysis close to the boundary of the form
  \begin{align}
 Z_1\sim r^{\Delta_1},\quad Z_2\sim r^{\Delta_2},\quad Z_3 \sim r^{\Delta_3}\,,
 \end{align}
 reveals modes with scaling dimensions $\Delta_1=3+z+\Delta_2, \Delta_3=3-z+\Delta_2$ where 
  \begin{align}
& \Delta_2=\{D-2+z, D-z,2-2z,\beta_\pm\}\,\nonumber\\
&\beta_\pm=\begin{cases}
 \frac{1}{2}\left(3+z\pm\sqrt{57+46z+9z^2}\right) \text{ for $D=5$} \,\nonumber\\
 \frac{1}{2}\left(2+z\pm\sqrt{20+28z+9z^2}\right) \text{ for $D=4$}\,.
 \end{cases}
 \end{align}
 
More concretely, the boundary conditions for the master fields take the form 

\begin{align}
Z_1&=c_1 r^{-D+2-z+\beta_-}+\tilde{c}_1 r^{-D+2-z+\beta_+}\nonumber\\
Z_2&= c_2 r^{D-2+z}+c_b\,c_1 r^{\beta_-}+c_d\,\tilde{c}_1 r^\beta_++c_4 r^{2-2z}\nonumber\\
Z_3&= c_a\,c_2 r^{2z}+c_3 r^2+c_c\,c_1 r^{\beta_-}+c_f\,\tilde{c}_1 r^\beta_++c_g\,c_4 r^{2-2z}
\end{align}
where $c_a,c_b,c_c,c_d,c_e,c_f,c_g$ are known functions of $z$ \textemdash\, the UV asymptotics of the field fluctuations are recorded in appendix \ref{app:UVsoundEMD}.  Given the above expansion, by setting $\tilde{c}_1,c_2,c_3=0$ we are demanding that the boundary sources vanish.

\begin{figure}[h!]
\centering
\includegraphics[width=0.50\linewidth]{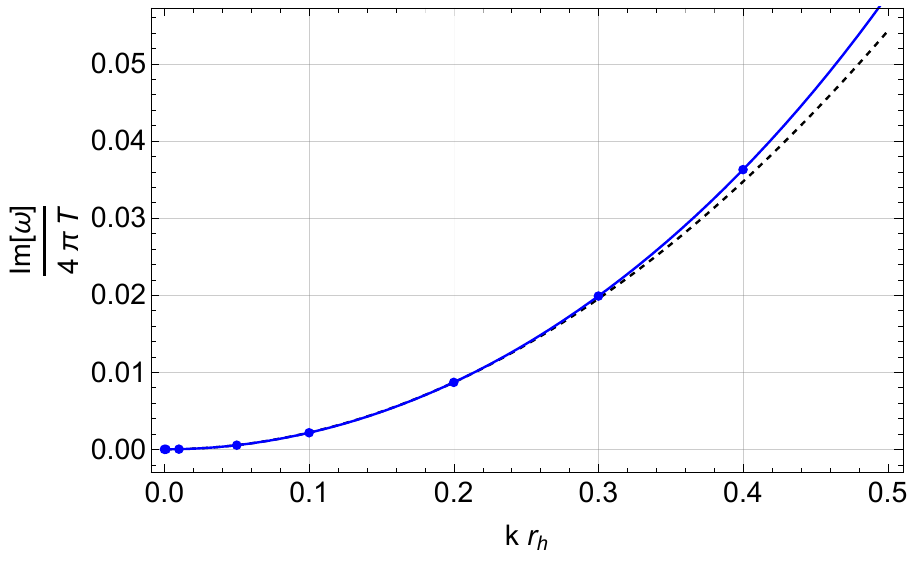}\quad\includegraphics[width=0.45\linewidth]{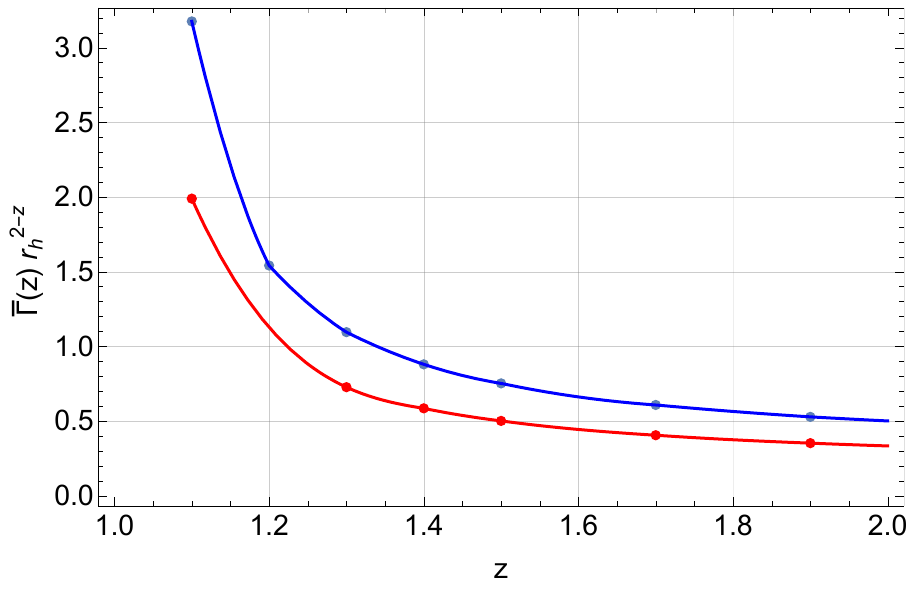}
\caption{Hydrodynamic modes for the sound channel in the EMD model. {\bf (a)} Plot of the dispersion relation as a function the momentum for $z=3/2$ for $D=4$. {\bf (b)} Values of the relaxation constant $\bar{\Gamma}(z)$ for $D=4$ (blue) and $D=5$ (red).}\label{fig:soundEMD}
\end{figure}

Imposing these boundary conditions, we solve the corresponding eigenvalue problem to find the following dispersion relation for the sound mode 
\begin{equation}
\omega_{sound}=-i \bar \Gamma(z) k^2+\dots \,,
\end{equation}
indicating that for the Einstein-Maxwell-Dilaton case the speed of sound is zero. In Fig. \ref{fig:soundEMD}(a), we plot the dispersion relation for $z=3/2$, $D=4$ and we see a clear quadratic scaling \textemdash \,this was also confirmed by studying the logarithmic derivative of the dispersion $k Im[\omega]'/Im[\omega]$. We find analogous behaviour for other values of $z$ and $D$. We depict  the behaviour of $\bar \Gamma(z)$ in Fig.  \ref{fig:soundEMD}(b). We confirmed these results by repeating the calculation using a gauge fixing term instead of master fields.

\subsection{Analytic results for $z=D-2$ in the vector channel}  

In this section we carry a perturbative analysis in an attempt to get an analytic handle on the numerical results we obtained for the shear channel. Due to the non-analytic behaviour of the metric functions, the hydrodynamic expansion can only be pushed analytically for $z = D-2$. 
 
In particular, we consider a perturbative expansion in the momentum $k$ of the form
\begin{equation}
k=\epsilon \,\bar k\,,\quad \omega=\sum_{i=1}^n \epsilon^i  \,\bar \omega^{(i)}
\end{equation}
 together with
\begin{align}
&\delta a_y= (1-u^\beta)^{-i \omega/4\pi T} \sum_{i=0}^n  \epsilon^i a_y^{(i)}(u)\,,\nonumber\\
&\delta H_{ty}= (1-u^\beta)^{-i \omega/4\pi T} u^3 \sum_{i=0}^n  \epsilon^i \,\delta H_{ty}^{(i)}(u)\,,
\end{align}
where $u=r_h/r$, $\beta=z+D-2$ and the prefactors were chosen in such a way to simplify the IR boundary conditions. Note that, in units of $r_h=1$, we have  $4\pi T=z+D-2$. At each order, $i$, we find two second order equations, which we can recast as a single higher order ODE for $\delta  a_{y}^{(i)}(u)$. Solving these equations, we find that at zeroth order
\begin{equation}
\delta a_y^{(0)}(u)=c_0,\quad \delta H_{ty}^{(0)}(u)=0\,,
\end{equation}
while at order 1 (and in fact for all odd orders)
\begin{equation}
\bar\omega^{(1)}=0\,,\quad \delta a_y^{(1)}(u)=0,\quad \delta H_{ty}^{(1)}(u)=0\,.
\end{equation}
Finally, we were able to solve the corresponding ODE at quadratic order and determine $ \bar\omega^{(2)}=0$ through imposing boundary conditions. Focusing on $D=4$, in the IR the solution looks like 
\begin{align}
&\delta  a_y^{(2)}(u)=
                        \frac{c_0}{192}{\bar k}^2(12-\pi^2)+c_2 +\frac{(1-u)}{32}c_0\,{\bar k}^2(12-\pi^2)+\dots\,,\nonumber\\
&\delta  H_{ty}^{(2)}(u)=
                         \frac{c_0 }{36} (-9+14{\bar k}^2 ) + \frac{c_0}{18} (9-8{\bar k}^2) (1-u) +\dots\,,
                     \end{align}
and in the UV
\begin{align}
&\delta  a_y^{(2)}(u)=
                        \frac{c_0\pi^2}{96}{\bar k}^2+c_2+\frac{c_1\,u^2}{72}(9-14{\bar k}^2+12 {\bar k}^2 \log u)+\dots\,,\\
 & \delta H_{ty}^{(2)}(u)=
                         - \frac{c_0}{36} u^2 (9-14{\bar k}^2 +12\,{\bar k}^2 \log u) +\dots\,.
\end{align}
Similar results apply for $D=5$. Determining analytic solutions to the higher order equations was not possible.

To summarise,  from this perturbative solution we have been able to conclude that the quadratic piece vanishes, $ \bar\omega^{(2)}=0$, 
which is consistent with our numerical results. Unfortunately we were not able to solve the equations to a sufficiently high order in perturbation theory to determine the value of the first non-trivial term, namely $ \bar\omega^{(4)}$.

\section{Comparing with Lifshitz hydrodynamics}\label{sec:FT}

The hydrodynamic modes of homogeneous and isotropic non-relativistic fluids with generic dynamical exponent $z$ was discussed in \cite{Hoyos:2013qna,Hoyos:2013eza,deBoer:2017abi}, with the main difference between the two approaches being that in \cite{Hoyos:2013qna,Hoyos:2013eza} boosts are broken at the first dissipative level while in \cite{deBoer:2017abi} they are already absent at the ideal level. In particular, \cite{deBoer:2017abi} derived the following hydrodynamic dispersion relations for sound and shear
\begin{align}
\omega_{sound}&=u_s k-i k^2 \Gamma+\dots\,,\nonumber\\
\omega_{shear}&=-i \frac{\eta_0}{\rho_0} k^2+\dots
\end{align}
 where $\rho$ is the mass density, $\eta$ is the charge density  and the zero index denotes equilibrium. They also expressed the speed of sound $u_s$ and the attenuation $\Gamma$ in terms of equilibrium quantities as follows
 \begin{align}
 &u_s^2=\frac{z}{d} \frac{\tilde \epsilon_0+p_0}{\rho_0}\,,\nonumber\\
 &\Gamma=\frac{1}{d}(d-1)\frac{\eta_0}{\rho_0} +\frac{\bar{\pi}_0}{2\rho_0} u_s^2
 \end{align}
where $d=D-1$, $\tilde \epsilon$ is the internal energy, $p$ is the pressure and $\bar{\pi}$ is the dissipative part of the thermal conductivity.

In the case of the Einstein-Proca model the dispersion relations we found numerically is consistent with the behaviour above for non-vanishing values of all parameters, in a way which smoothly connects to the AdS values as $z \to 1$. 
However, for the EMD model we numerically find that the speed of sound as well as the leading term in the shear dispersion relation vanish. The latter was also confirmed analytically for a particular choice of $z$. To reconcile the two results, $\rho_0\to\infty$ together with $\bar{\pi}_0\sim \rho_0^2$ while the rest of the coefficients are order one. Checking this explicitly is left for future work.
 
\section{Discussion and Future Direction}\label{sec:discussion}

In this work we have considered linearised perturbations around Lifshitz black branes, with a focus on electromagnetic and gravitational fluctuations in the scalar and vector channels. We found that in the case of the Einstein-Proca model, the dispersion relations of the hydrodynamic shear and sound mode have the same structure as in asymptotically $AdS$ black branes, namely they have quadratic and linear dispersions, respectively. 
On the other hand, we found that the dispersions of the shear and sound modes in Einstein-Maxwell-Dilaton model are quartic and quadratic, respectively. This is a significant difference between the two models, signalling that the two models approach equilibrium differently. It would be interesting to understand this better from the field theory perspective. As mentioned in Sec. \ref{sec:EMD}, the $z \to 1$ limit of the EMD system does not smoothly approach the relativistic Einstein-Maxwell theory, at least taken while keeping all other quantities regular in $(z-1)$. This could explain the root of the qualitative change we observe in the dispersion relations of the hydrodynamic modes. 
 
Preliminary analysis of the near-horizon expansion of the perturbation indicates the phenomenon of pole-skipping does occur in asymptotically Lifshitz geometries. Specifically, we find that the Matsubara frequencies exist in the lower half plane at the exact same locations as in the relativistic case. It would be interesting to check if the Lifshitz hydrodynamic sound mode, when driven to instability by a choice of a specific value of imaginary momentum that is well outside the standard regime of validity of hydrodynamics, exhibits connections with chaos through the phenomenon of pole-skipping \cite{Grozdanov:2017ajz}.  In addition, in the spirit of \cite{Withers:2018srf,Jansen:2020hfd}, it would be interesting to further investigate the radius of convergence of the non-relativistic hydrodynamic expansion and, through resummation, test whether it is possible to extract non-hydrodynamic QNMs from the hydrodynamic ones. This will be shed light on the thermalisation properties of strongly coupled non-relativistic fluids \textemdash  for relativistic fluid thermalisation, it is known that the hydrodynamic expansion becomes applicable very early during dynamical processes. 

An exciting recent technical development with the potential of leading to rapid progress in both gravitational physics and holography involves the limit of large number of dimensions ($D$) of general relativity \cite{Emparan:2015rva,Emparan:2014cia}. This tool has so far been applied to relativistic theories of gravity, realised on geometries that are either asymptotically flat or asymptotically AdS, providing a number of very interesting results. By treating $D$ as a free parameter, one can use this approach to perform a perturbative expansion in $1/D$, which leads to a drastic simplification of the theory. The non-trivial black hole dynamics are localised within a distance $1/D$ from the horizon and it is thus possible to capture them with an effective theory given by a set of constraints that depend solely on the directions parallel to the horizon. It would be interesting to explore the applicability of this tool in spacetimes with Lifshitz asymptotics. The first step in this direction would be to see if the tower of gravitational QNMs splits into two subgroups, one controlling the dynamics in the near horizon region and the other in the far region, when $D\to\infty$.

\section*{Acknowledgements}
We would like to thank Andrei Starinets for collaboration in early stages of the project, and Watse Sybesma and Stefan Vandoren for useful discussions. T.A. is supported in part by the ERC Advanced Grant GravBHs-692951 and by Grant CEX2019-000918-M funded by Ministerio de Ciencia e Innovación (MCIN) / Agencia Estatal de Investigación (AEI) / 10.13039 / 501100011033. 
C.P. acknowledges support by the European Union’s Horizon 2020 research and innovation programme under the Marie Sklodowska-Curie grant agreement HoloLif No 838644 and from a Royal Society - Science Foundation Ireland University Research Fellowship via grant URF/R1/211027.

\appendix

\section{Asymptotics for the sound fluctuations in the Einstein-Proca model}\label{app:EP}

We specify ingoing boundary conditions for all the fluctuations, which translates to demanding the following behaviour near the horizon
 \begin{align}
& \delta g_{tt}=r^{2z} F(r)(r-r_h)^{-i \omega/4\pi T-1} \delta g_{tt}^{(reg)}\,,\nonumber\\
&  \delta g_{tr}=(r-r_h)^{-i \omega/4\pi T-1}\delta g_{tr}^{(reg)}\,,\nonumber\\
&  \delta g_{tx}=(r-r_h)^{-i \omega/4\pi T}\delta g_{tx}^{(reg)}\,,\nonumber\\
&  \delta g_{rr}=\frac{r^2}{R(r)}(r-r_h)^{-i \omega/4\pi T-1}\delta g_{rr}^{(reg)}\,,\nonumber\\
&   \delta g_{rx}=(r-r_h)^{-i \omega/4\pi T-1}\delta g_{rx}^{(reg)}\,,\nonumber\\
&   \delta g_{xx}=r^2(r-r_h)^{-i \omega/4\pi T}\delta g_{xx}^{(reg)}\,,\nonumber\\
&\delta g=r^2(r-r_h)^{-i \omega/4\pi T}\delta g^{(reg)}\,,\nonumber\\
& \delta A_{t}=(r-r_h)^{-i \omega/4\pi T}\delta a_{t}^{(reg)}\,,\nonumber\\
& \delta A_{r}=(r-r_h)^{-i \omega/4\pi T-1}\delta a_{r}^{(reg)}\,,\nonumber\\
&\delta A_{x}=(r-r_h)^{-i \omega/4\pi T}\delta a_{x}^{(reg)}\,,
 \end{align}
where $(reg)$ indicates that the respective functions admit regular power series expansions in the near horizon. On the other hand, close to the boundary we get an expansion of the form \cite{Andrade:2013wsa}
  \begin{align}\label{eq:fieldsUVexpEP}
\delta g_{tt}=&r^{2z} F(r)\left( g_{tt}^{(0)}+  g_{tt}^{(v)} r^{-z-D+2}+s^{(-)}_1\,r^{\beta_-}+s^{(-)}_2\,r^{\gamma_-}+s^{(+)}_1\,r^{\beta_+}+s^{(+)}_2\,r^{\gamma_+}\dots\right)\,,\nonumber\\
\delta g_{tr}=&g_{tr}^{(0)} r^{2z-1}+g_{tr}^{(v)}r^{-z-D+1}+\dots\,,\nonumber\\
\delta g_{tx}=&r^{2z}\left(g_{tx}^{(0)}+g_{tx}^{(v)}r^{-D+4-3z}+g_{tx}^{(c)}r^{-z-D+2}+\dots\right)\nonumber\\
	             &+r^{2}\left(\bar{g}_{tx}^{(0)}+\bar{g}_{tx}^{(v)}r^{z-D}+\bar{g}_{tx}^{(c)} r^{-z-D+3}+\dots\right) \,,\nonumber\\
\delta g_{rr}= &\frac{r^2}{R(r)}\left(g_{rr}^{(0)}+ g_{rr}^{(v)}r^{-z-D+2}+s^{(-)}_1\,r^{\beta_-}+s^{(-)}_2\,r^{\gamma_-}+s^{(+)}_1\,r^{\beta_+}+s^{(+)}_2\,r^{\gamma_+}\dots\right)\,,\nonumber\\
\delta g_{rx}=&g_{rx}^{(0)}r^{1}+g_{rx}^{(v)}r^{-z-D+1}+s^{(-)}_1\,r^{\beta_-}+s^{(-)}_2\,r^{\gamma_-}+s^{(+)}_1\,r^{\beta_+}+s^{(+)}_2\,r^{\gamma_+}+\dots \,,\nonumber\\
\delta g_{xx}=&r^2\,\left( g_{xx}^{(0)}+g_{xx}^{(v)} r^{-z-D+2}+s^{(-)}_1\,r^{\beta_-}+s^{(-)}_2\,r^{\gamma_-}+s^{(+)}_1\,r^{\beta_+}+s^{(+)}_2\,r^{\gamma_+}\dots\right)\,,\nonumber\\
\delta g=&r^2\,\left( g_{yy}^{(0)}+ g_{yy}^{(v)}r^{-z-D+2}+s^{(-)}_1\,r^{\beta_-}+s^{(-)}_2\,r^{\gamma_-}+s^{(+)}_1\,r^{\beta_+}+s^{(+)}_2\,r^{\gamma_+}\dots\right)\,,\nonumber\\
\delta A_{t}=& r^z\left(a_{t}^{(0)}+ a_{t}^{(v)}r^{-z-D+2}+s^{(-)}_1\,r^{\beta_-}+s^{(-)}_2\,r^{\gamma_-}+s^{(+)}_1\,r^{\beta_+}+s^{(+)}_2\,r^{\gamma_+}\dots\right)\,,\nonumber\\
 \delta A_{r}=&a_{r}^{(0)}r^{z-1}+a_{r}^{(v)}r^{-D+1-2z}+\dots \,,\nonumber\\
\delta A_{x}= &r^z\left(a_{x}^{(0)}+a_{x}^{(v)}r^{-D+4-3z}+a_{x}^{(c)}r^{-z-D+2}+\dots\right)\,,
 \end{align}
 where
 \begin{align}
 &\gamma_\pm=\begin{cases}
 				-\frac{1}{2}(z+4\pm\sqrt{9z^2+4z+20})\qquad \text{for D=4}\nonumber\\
				-\frac{1}{2}(z+5\pm\sqrt{9z^2+6z+33})\qquad \text{for D=5}\,,\nonumber\\
				\end{cases}\\
 &\beta_\pm=\begin{cases}
 			-\frac{1}{2}(z+4\pm\sqrt{9z^2-20z+20})\qquad \text{for D=4}\nonumber\\
			-\frac{1}{2}(z+5\pm\sqrt{9z^2-26z+33})\qquad \text{for D=5}\,.
			\end{cases}
 \end{align}
Note that \eqref{eq:fieldsUVexpEP} is meant to capture the general form of the expansion and not all coefficients are independent. Furthermore, just like in the DeTurk trick, the non-analytic terms, $\gamma_\pm$, are an artefact of the way we fix the gauge and the free constants that multiply it are in fact zero on the actual solution. In this expansion we identify the leading coefficients $g_{tt}^{(0)}$, $g_{tx}^{(0)}$, $\bar{g}_{tx}^{(0)}$, $g_{rr}^{(0)}$, $g_{xx}^{(0)}$, $g_{yy}^{(0)}$, $a_{t}^{(0)}$, $a_{x}^{(0)}$,$s_i^{(-)}$ as sources and we demand that they vanish.

\section{Asymptotics for the sound fluctuations in the Einstein-Maxwell-Dilaton model}
\label{app:UVsoundEMD}

The UV expansions for the field fluctuations in the scalar sector of the EMD model take the form
 \begin{align}\label{eq:fieldsUVexpEMD}
 \delta g_{tt}=&r^{2z} f(r)\left(g_{tt}^{(0)}+g_{tt}^{(v)}r^{2-D-z}+\dots\right)\,,\nonumber\\
 \delta g_{tx}=&r^{2z}\left(g_{tx}^{(0)}+g_{tx}^{(v)} r^{-D+4-3z}+g_{tx}^{(c)} r^{-z-D+2}+\dots\right)\nonumber\\
		     &+r^{2}\left(\bar{g}_{tx}^{(0)}+g_{tx}^{(v)} r^{z-D}+{g_{tx}}^{(c)} r^{-z-D+2}+\dots\right)\,,\nonumber\\
 \delta g_{xx}=&r^2\left(g_{xx}^{(0)}+g_{xx}^{(v)}r^{2-D-z}+\dots\right)\,,\nonumber\\
 \delta g=&r^2\left(g_{yy}^{(0)}+g_{yy}^{(v)}r^{2-D-z}+\dots\right)\,,\nonumber\\
 \delta A_{t}=&r^{z+D-2}\left(a_t^{(0)}+a_t^{(v)} r^{-D+4-3z}+a_t^{(c)} r^{-z-D+2}+\dots\right)\,,\nonumber\\
 \delta A_{x}=&r^{z+D-2}\left(a_x^{(0)}+a_x^{(v)} r^{-D+4-3z}+a_x^{(c)} r^{-z-D+2}+\dots\right)\,,\nonumber\\
 \delta \phi=&\phi^{(0)} r^{\beta_-}+\phi^{(d)}+\phi^{(v)}r^{\beta_+}+\phi^{(c)}r^{-2+z}+\dots\,,
 \end{align}
which, at least for the majority of the metric fields, looks similar to \eqref{eq:fieldsUVexpEP}. Note that \eqref{eq:fieldsUVexpEMD} is meant to capture the general form of the expansion and not all coefficients are independent. Close to the UV boundary we impose 
boundary conditions that kill the leading modes $g_{tt}^{(0)}$, $g_{tx}^{(0)}$, $\bar{g}_{tx}^{(0)}$, $g_{xx}^{(0)}$, $g_{yy}^{(0)}$, $a_{t}^{(0)}$, $a_{x}^{(0)}$, $\phi^{(0)}$ as they correspond to boundary sources.

\bibliographystyle{JHEP-2}
\bibliography{Lifz}{}
\end{document}